\documentclass[preprintnumbers,amsmath,amssymb]{revtex4}
\usepackage{graphicx}
\usepackage{amsmath,amssymb,latexsym}

\begin{document}

\title{Anomalous scaling due to correlations: Limit theorems and self-similar processes}
 
\author{Attilio L. Stella}
\email{stella@pd.infn.it}
\affiliation{
Dipartimento di Fisica,
Sezione INFN and CNISM, Universit\`a di Padova,\\
\it Via Marzolo 8, I-35131 Padova, Italy
}

\author{Fulvio Baldovin}
\email{baldovin@pd.infn.it}
\affiliation{
Dipartimento di Fisica,
Sezione INFN and CNISM, Universit\`a di Padova,\\
\it Via Marzolo 8, I-35131 Padova, Italy
}

\date{\today}

\begin{abstract}
We derive theorems which outline explicit mechanisms by
which anomalous scaling for the probability density function of the 
sum of many correlated random variables asymptotically prevails. 
The results characterize general  
anomalous scaling forms, justify their universal character, and specify
universality domains in the spaces of joint
probability density  functions of the summand variables. 
These density functions are assumed to be invariant under 
arbitrary permutations of their arguments. Examples from the
theory of critical phenomena are discussed.
The novel notion of stability implied by the limit theorems
also allows us to define sequences of random variables whose sum 
satisfies anomalous scaling for any finite number of summands. If regarded 
as developing in time, the stochastic processes
described by these variables are non-Markovian generalizations of 
Gaussian processes with uncorrelated increments, and provide, e.g.,  
explicit realizations of a recently proposed model of 
index evolution in finance. 
\end{abstract}

\maketitle

\section{Introduction} 
A major achievement of the theory of probability are the limit 
theorems \cite{gnedenko_1,feller_1}, 
which provide the basis to explain statistical
regularities observed in large classes of natural, economical and 
social mass-scale phenomena. 
These theorems describe the mechanisms leading to universal forms of 
scaling for the probability density functions (PDF's) of 
sums of many independent random variables.
The scaling can be normal, or anomalous, depending on
whether the PDF's of the individual 
variables possess finite second moment, or not.
However, independence is not guaranteed in general, and
a large number of collective phenomena in Nature exhibit anomalous scaling
\cite{kadanoff_1,kadanoff_2,sethna_1,bouchaud_3,lu_1,scholz_1,kiyono_1,
lasinio_1,lasinio_2,vulpiani_1,eisler_1,clusel_1,hilhorst_1}
as a consequence of correlations. 
In such cases, 
if the PDF of the sum of the
elementary variables and its argument are simultaneously rescaled by a power $D$
of the number of summands, it asymptotically converges to 
a scaling function $g$ which is not necessarily  Gaussian nor
L\'evy, and the scaling exponent $D$ is in general not equal to
$1/2$. 
Thus, an  open challenge 
remains that of establishing limit theorems able to justify the 
existence and the 
universality of the anomalous scaling forms occurring in the case 
of strongly correlated variables.

The renormalization group approach to critical phenomena in
statistical physics \cite{kadanoff_1} has led to developments 
in probability theory which point towards a solution of this problem.
Indeed, the fixed-point condition for block-spin transformations can 
be regarded \cite{lasinio_1,lasinio_2,vulpiani_1} as a substitute of the stability condition 
at the basis of the limit theorems for the independent 
case \cite{gnedenko_1,feller_1}.
For instance, in the context of hierarchical equilibrium spin 
models the fixed-points of these block-spin transformations 
are expected to attract whole domains of strongly correlated critical 
systems displaying asymptotically the same universal form of anomalous 
scaling \cite{lasinio_1,lasinio_2,vulpiani_1}. However, unlike in the case of the limit 
theorems for independent variables, classes of admissible universal
scaling forms and their universality domains are not easily
identified.  

Since the standard limit theorems hold in force of the  
multiplicative structure of the joint PDF's 
of independent variables, an attempt has been recently
made by the present authors \cite{baldovin_1} 
to establish theorems on the basis of a generalization 
of the multiplication operation, leading to dependent joint
probability densities. Yet, due to mathematical difficulties, 
the problem of constructing consistent joint PDF's for
correlated variables whose sum asymptotically 
satisfies scaling was not addressed \cite{baldovin_1}.  

Correlated random variables often considered in probability 
theory are those in exchangeable sequences \cite{aldous_1}. 
The joint PDF's of an arbitrary number of variables
in an exchangeable sequence have the property of being invariant 
under permutations of their arguments.
Exchangeability was introduced by
de Finetti \cite{definetti_1}, and is of paramount importance in the 
Bayesian approach to probability and statistics \cite{aldous_1}. 
It is already known that, thanks to the simplifying feature of
exchangeability, central limit theorems can be
established \cite{jiang_1}. 
The scalings foreseen by these theorems for the PDF of the 
sum of the random variables 
involve scaling functions which are 
convex combinations (mixtures) of Gaussians.
For $D$ only two values could be considered. 
If the variables are linearly uncorrelated, i.e., correlations
are nonzero only for nonlinear functions of the variables,
the scaling exponent is the ordinary $D=1/2$ \cite{jiang_1}. 
Alternatively, if the variables are correlated also at linear level, 
limit theorems have been proved for $D=1$ \cite{teicher_1}. 

Inspired by ideas from the modern theory of critical phenomena, 
in the present Article we establish limit theorems
for sums of $N$ dependent random variables whose joint PDF's,
upon increasing $N$, do not define sequences of random variables,
in general. With those defining exchangeable variable sequences, 
our joint PDF's only share the property of being invariant
under permutations of their arguments. To illustrate how PDF's with 
such properties can arise in physics, we discuss the example
of a permutationally invariant description of a magnetic system.
The novel theorems apply to anomalous scalings with general
exponent $D$. They also enable the 
explicit construction of universality domains, i.e. of whole classes
of sequences of joint PDF's sharing asymptotically the
same scaling form for the sum of the variables.

The limit theorems proved here have implications also
outside the context of variables with permutationally invariant
joint PDF's. Indeed, they were inspired by a recent proposal
for the description of the time evolutions of financial indexes as
stochastic processes \cite{baldovin_2,stella_1}. 
When dealing with such processes, one often
considers time series in which each term represents the increment 
of an additive collective variable in an elementary time
interval. Examples are the displacement in diffusion,
or the logarithmic return of a financial asset. 
In these cases, causality imposes that the successive
increments must constitute a sequence of random variables, in which 
the statistical properties of each variable are independent of the 
successive ones. 
When the increments are correlated
and the processes have the property of self-similarity, i.e. when the collective variable 
distribution obeys scaling not just asymptotically, but for any finite number of summands,
there are some requirements whose satisfaction
has to be imposed to the joint PDF's
of the successive increments. An heuristic way of 
satisfying these requirements was recently proposed as a basis
for a stochastic model of the dynamics of financial indexes
\cite{baldovin_2,stella_1}.
As we show in this work, the heuristic proposal in \cite{baldovin_2,stella_1}
is fully justified on the basis of the novel notion of
stability implied by our theorems. 

In general our stochastic processes are non-stationary
and the scaling has a time-inhomogeneous nature \cite{baldovin_3}.
When they become stationary, their increments also constitute 
sequences of exchangeable random variables. In such cases it is not
possible to reproduce the statistics of these variables
by empirical time-averages along infinitely long, single realizations of
the processes.
This is due to a mechanism of {\it ergodicity breaking}
implied by de Finetti's representation theorem
\cite{aldous_1,definetti_1}. A way out of this difficulty is found when
considering self-similarity as a property of the process valid
within a limited, although possibly large, range of time-scales. 
This attitude is fully legitimate in many applications \cite{kadanoff_3}. 
We show here, by a dynamical 
simulation strategy of wide use in finance \cite{engle_1},
how ergodicity can be restored in the process,
by requiring scale-invariance to hold only up to a finite upper
cutoff in time.

This Article is organized as follows. In the next three Sections, we
introduce the formalism and 
present our main results about the limit theorems. 
We enunciate these theorems and give full details of their derivations 
in the Appendix. After stressing the applicability
of our approach to the forms of anomalous scaling
emerging, e.g., in the context of critical phenomena, 
in Sections \ref{section_nm_ss_processes} and \ref{section_restoring}
we discuss implications of our results for 
the theory of stochastic processes. In particular, we
present a class of non-Markovian self-similar processes
possessing  the requisites recently postulated
\cite{baldovin_2,stella_1} for the case of finance and allowing explicit
analytical calculations and efficient simulation strategies.
The last Section is devoted to conclusions.

\section{Anomalous scaling}

Let us consider, for any given $N=1,2,3,\dots$, a set of random
variables, $X_i$, with $i=1,2,\dots,N$, taking values $x_i$ on
the real axis.
We call $p_{N}(x_1,\dots ,x_N)$ the joint
PDF of $N$-th set of variables and, to start with, assume that for any $N$ this 
function is invariant under arbitrary permutations of its arguments.
It should be stressed that, e.g., the random variable $X_1$ belonging
to a set with $N$ variables and the $X_1$ belonging to another
set with $N'\neq N$ variables are not identical, in general.
Thus, in principle we should denote the variables in the $N$-th set
by $X_i^{(N)}$, $i=1,2,\dots,N$, and their values by $x_i^{(N)}$.
However, in order to keep formulas simple, we will not adopt this
notation. Ultimately the identity of each
variable $X_i$ will be specified by
the joint PDF $p_N(x_1,x_2,\dots,x_i,\dots,x_N)$ used in 
order to evaluate its statistical properties.
In this way, our formulas will conform to the standards 
of the statistical mechanics literature \cite{lasinio_1,lasinio_2,hilhorst_1}.
To further simplify the formalism we can require, without loss of
generality, that for any $N$ all the variables 
have zero average,
$\langle X_i\rangle_{p_N}=0$ $\forall i$,
where $\langle (\cdot)\rangle_{p_N}\equiv
\int d x_1\cdots d x_N \;(\cdot)\;p_N(x_1,\ldots,x_N)$.
For the sum $Y_N\equiv X_1+\dots +X_N$, whose PDF is
\begin{equation}
p_{Y_N}(y)=\int d x_1\cdots d x_N \;\delta(y-x_1-\ldots-x_N)\;p_N(x_1,\ldots,x_N),
\end{equation}
this also implies $\langle Y_N\rangle_{p_{Y_N}}=0$.
We are interested in cases in which the sequence
$p_N(x_1,\dots,x_N)$, $N=1,2,\dots$ is such that $p_{Y_N}$ satisfies
anomalous scaling for $N\to \infty$, i.e.
\begin{equation}
N^D\:p_{Y_N}(N^Dy)\to g(y), 
\label{eq_scaling}
\end{equation}
where $g$ is
a scaling function, and $D$ is
a scaling dimension. 
We want to identify whole domains
of $p_{N}$'s such that the $p_{Y_N}$
satisfies Eq. (\ref{eq_scaling})
with a given $g$ and a given $D$.
Besides the kind of convergence, the class of admissible 
$g$'s and the range of $D$'s needs to be specified.   
As we discuss below, examples of $p_{Y_N}$'s such that
Eq.(\ref{eq_scaling}) holds are easily found in statistical physics.

We first clarify why the exponent values $D=1/2$ and $D=1$
naturally arise for sequences of exchangeable variables. 
Let us suppose that $\langle Y_N^2\rangle_{p_N}$ is finite for any $N$.
Since permutational invariance implies  
$\langle X_i\rangle_{p_N}=\langle X_1\rangle_{p_N}$ $\forall i$, 
and 
$\langle X_i X_j\rangle_{p_N}=\langle X_1X_2\rangle_{p_N}$ $\forall i\neq j$,
one has
\begin{equation}
\langle Y_N^2\rangle_{p_N}=N \langle X_1^2\rangle_{p_N}+
N(N-1)\langle X_1X_2\rangle_{p_N}.
\label{eq_2nd_mom}
\end{equation}
On the other hand, if, as appropriate for sequences of random variables,
the sequence of joint PDF's $p_N$ is constructed consistently with the 
condition
\begin{equation}
p_{N-1}(x_1,\ldots,x_{N-1})=
\int d x_N p_N(x_1,\ldots,x_N),
\label{eq_reduced}
\end{equation}
where $N \geq 2$, it is clear that $\langle X_1\rangle_{p_N}$
and $\langle X_1X_2\rangle_{p_N}$ do not depend on $N$.
Since according to the scaling condition in Eq. (\ref{eq_scaling}) 
$\langle Y_N^2\rangle_{p_N}\sim N^{2D}$, Eq. (\ref{eq_2nd_mom}) implies 
that either $D=1/2$ and $\langle X_1X_2\rangle_{p_N}=0$, or  
$D=1$ and $\langle X_1X_2\rangle_{p_N}>0$. In the former case, 
further restrictions on the averages of products of $X$'s
apply if higher moments of $Y_N$ are assumed to exist.
We should stress that if Eq.(\ref{eq_reduced}) is satisfied by the sequence
of permutation-invariant joint PDF's, then these PDF's in turn define a sequence
of exchangeable variables. Indeed, Eq.(\ref{eq_reduced}) guarantees that a
given variable, say $X_1$, is strictly the same random variable, 
independent of the set of $N$ variables within which it is considered.

As discussed in Section \ref{section_permutation}, there are cases, for 
example in statistical mechanics, where one
considers a system in equilibrium at a given temperature,
so that $p_N$ represents the canonical
joint PDF of $N$ variables describing the degrees of freedom of the 
system. Since $p_N$ is expressed as a ratio between the Gibbsian weight
and the partition sum, upon integrating $p_N$ over one of the $N$ 
variables, as a rule we do not obtain the joint PDF of a system in 
equilibrium at the same temperature and with just $N-1$ variables.
Indeed, tracing over one of 
the variables leads to effective interactions which are not present in 
the Hamiltonian for $N-1$ variables.
The modern theory of critical phenomena shows that the renormalization 
effects determining this difference lead to anomalous scaling at the 
critical point \cite{kadanoff_1}.
This circumstance, which is expected to occur in many cooperative
phenomena, will allow us to derive limit theorems for sums of
exchangeable variables with general values of $D$.

On the other hand, 
in problems where $N$ represents the number of increments
over successive time intervals of a stochastic process
and $p_{N-1}$ and $p_N$ are respectively 
the joint PDF's of the first $N-1$ and $N$ increments, causality
imposes to consider sequences of $p_N$'s satisfying Eq. (\ref{eq_reduced}). 
Below we will also show how the stability conditions implied by our
limit theorems allow to define sequences of random variables
whose joint PDF's satisfy Eq.(\ref{eq_reduced}) and whose aggregated increment $Y_N$
satisfies anomalous scaling exactly for any $N$.

\section{Illustration of the main results}
We report our main statements and their mathematical proofs in the 
Appendix. Here we rather choose to illustrate the meaning 
and some implications of our results. Let us first consider $p_N$ of the 
form
\begin{equation}
p_N(x_1,x_2,\dots,x_N) = \int_{-\infty}^{+\infty} d\mu\;
\lambda(\mu)\;\prod_{i=1}^N\;
l\left(x_i-\frac{\mu}{N^{1-D}}\right),
\label{eq_joint_1}
\end{equation}
where $\lambda$ and $l$ are single-variable PDF's.
With no loss of generality we require $\langle\mu\rangle_\lambda=0$, 
whereas we assume
$\langle X\rangle_l=0$ and $\langle X^2\rangle_l=1$. 
The higher integer moments of $l$, are left arbitrary.
Clearly, the $p_N$ in Eq. (\ref{eq_joint_1}) is 
a positive density normalized to $1$, and invariant under permutations of its 
arguments. The $X_i$'s are dependent, since $p_N$
does not simply factorize into a product of single variable PDF's. 
The choice of considering $p_N$'s which are convex combinations of products 
of single-variable PDF's is motivated by the fact that in this way it
is possible to demonstrate the existence of asymptotic scalings with
very general scaling functions.
Indeed, in the Appendix  we show that with the 
joint PDF in Eq. (\ref{eq_joint_1}), $p_{Y_N}$ satisfies 
Eq. (\ref{eq_scaling})
with a scaling exponent $D \geq 1/2$.
The scaling function $g$ is determined by $\lambda$.
For $D=1/2$ $g$ is given by
\begin{equation}
g(x)=\int_{-\infty}^{+\infty} d\mu\;\lambda(\mu)\;
\frac{\exp\left[-(x-\mu)^2/2\right]}{\sqrt{2\pi}},
\label{eq_g_lambda}
\end{equation}
whereas $g$ coincides with $\lambda$ itself if $D>1/2$. 
In both cases, upon varying $\lambda$ the scaling function  
$g$ assumes general shapes. For instance, it may have 
several local and global maxima and power law decays to zero at 
large positive $x$, and/or $-x$, as required in many applications. 

As anticipated above, when
the variables $X_i$'s are dependent and do not constitute a sequence, it 
is legitimate to introduce in the definition of $p_N$ the $N$-dependence
arising from the fact that $\mu$ enters divided by $N^{1-D}$. 
In particular, precisely this dependence  implies that
the joint PDF of a system with $N-1$ variables, $p_{N-1}$, 
rather than satisfying Eq. (\ref{eq_reduced}), is linked to $p_N$ by
the relation:
\begin{equation}
p_{N-1}(x_1,\ldots,x_{N-1})=
\left(\frac{N-1}{N}\right)^{(N-1)(1-D)}
\int d x_N\,p_N\left(
\left(\frac{N-1}{N}\right)^{(1-D)}x_1,\ldots,\left(\frac{N-1}{N}\right)^{(1-D)}x_{N-1},x_N
\right).
\label{eq_reduced_rg_1}
\end{equation}

Consistently with the fact that the $X_i$'s are not constituting a sequence
of random variables, the marginal PDF of each individual $X_i$, 
\begin{equation}
p_{X_i,N}(x_i)\equiv\int d x_1\cdots d x_{i-1}\,d x_{i+1}\cdots d x_N \,
p_{N}(x_1,\ldots,x_N),
\end{equation}
depends clearly on $N$ ($N \geq i$).
So, if the second moment of $\lambda$ is finite, one
realizes that $p_{X_i,N}$
has a finite width for $N \to \infty$ when $1/2\leq D\leq1$.
If $D>1$, this
width diverges in the large $N$ limit. Such a divergence makes full
sense in a correlated context. Indeed, in relation to the anomalous character
of the scaling, the marginal single-variable
PDF's play here a role analogous to that of single-variable PDF's
in the independent case. For example, 
with independent variables one allows the single
variable PDF's
to be of infinite width for any $N$, in order to have an
anomalous, L\'evy scaling limit \cite{gnedenko_1,feller_1} of $p_{Y_N}$. 
Here, with correlated variables, the dependence on $N$ 
entering in $p_{X_i,N}$ and the consequent divergence 
of width for $N \to \infty$ and $D>1$ play a qualitatively similar role 
in producing anomalous scaling. 

It is natural to ask what are the correlations of the
variables $X_i$'s according to the joint PDF's defined in
Eq. (\ref{eq_joint_1}). 
If $\langle\mu^2\rangle_{\lambda}$ exists, an easy calculation gives for example
\begin{equation}
\langle X_i X_j\rangle_{p_N} =\frac{\langle\mu^2\rangle_{\lambda}}{N^{2-2D}}
\label{eq_lin_corr_1}
\end{equation}
for $i \neq j$. 
In particular,
the variables with permutation-invariant joint PDF's as in Eq. (\ref{eq_joint_1}) 
are linearly correlated for finite $N$. 
When $1/2\leq D<1$ their linear correlators 
approach zero only asymptotically.

Next, we consider more general scaling functions which can be
expressed as convex combinations of Gaussians with varying
centers $\mu$ and widths $\sigma$. The form is
\begin{equation}
g(x)= \int_0^{+\infty} d\sigma \int_{-\infty}^{+\infty} d\mu\,\psi(\sigma,\mu)\,
\frac{\exp\left[-(x-\mu)^2/{2\sigma^2}\right]}{\sqrt{2\pi \sigma^2}},
\label{eq_g_psi}
\end{equation}
where $\sigma\in(0,\infty)$, 
and $\psi$ is a PDF. The scaling exponent can be now any 
$D>0$.
Again, for the sake of simplicity we require
$\langle\mu\rangle_\psi=0$,
while $\psi$ must be strictly equal to zero in a whole 
neighborhood of $\sigma=0$, for any $\mu$. 
In the Appendix we prove that with the $p_{N}$'s constructed
as follows: 
\begin{equation}
p_N(x_1,\dots,x_N)=
\int_0^{\infty} d\sigma \int_{-\infty}^{+\infty} d\mu\;
\psi(\sigma,\mu)\,\prod_{i=1}^N\,
\frac{l(x_i/\sigma N^{D-1/2}-\mu/\sigma N^{1/2})}{\sigma N^{D-1/2}},
\label{eq_joint_psi}
\end{equation}
with $\langle X\rangle_l=0$ and $\langle X^2\rangle_l=1$ as before,
$p_{Y_N}$ satisfies the asymptotic scaling
(\ref{eq_scaling}) with $g$ given by Eq. (\ref{eq_g_psi}) and the
chosen $D>0$.
One easily verifies that
if $\psi(\sigma,\mu)=\rho(\sigma)\,\delta (\mu)$
the $X$ 
variables are linearly uncorrelated for any $N$.
In this case, with $D=1/2$ the scaling limit of our 
theorem recovers known results valid for sums of random
variables in exchangeable sequences
\cite{jiang_1}.

It should also be noticed that if we put $\Psi(\sigma,\mu)=\delta(\sigma- 1)
\lambda(\mu)$ and $D=1/2$, one recovers the case discussed 
at the beginning of this section.

\section{Permutation-invariant joint PDF's and critical 
phenomena}
\label{section_permutation}
All the cases discussed in the previous sections concern
correlated variables whose joint PDF's for any $N$ are permutationally
invariant. At first sight, such feature may appear a too restrictive 
condition to be satisfied by realistic models, and
applications may often require to release it. 
However, in the study of anomalous scaling 
variables of this kind  may still play an important role. To illustrate 
this point, we consider the example of an Ising-like spin model, of the
type often studied in the renormalization
group approach to critical phenomena \cite{kadanoff_1}. 
Let us consider a system
of $N$ spins $S_i$, $i=1\dots,N$, where the index $i$ labels
the sites of a finite box of square or cubic lattice. The
spins are supposed to take values $s_i$ on the real axis.
Equilibrium statistical mechanics allows in principle to
construct the joint PDF of the $N$ spin variables once
given the spin Hamiltonian $H(\{ s \})$ and the temperature
$T$. 
Since the spin variables are associated to 
the lattice sites, their joint PDF is not invariant under
permutations.
Indeed, for any configuration $\{s_1,s_2,\dots,s_N\}$,
one has in general
$H(s_{\pi(1)},\ldots,s_{\pi(N)}) \neq
H(s_1,\ldots,s_N),
$
if $\pi$ is a permutation of the $N$ labels. 
This inequality holds because $H$ is a sum of local 
interactions. Thus, also the canonical joint PDF
\begin{equation}
p^\prime_N(s_1,\ldots,s_N)\equiv\frac{\exp[-H(s_1,\ldots,s_N)/k_B\,T]}
{\int \prod_{i=1}^N d s_i^\prime \exp[-H(\{s^\prime\})/k_B\,T]}
\end{equation}
where $k_B$ is the Boltzmann constant,
is not invariant under permutations of its arguments.
On the other hand, when, e.g., discussing the critical behavior of
the model, a key collective random quantity to be considered 
is the sum of all the spins $\sum_{i=1}^N S_i$ \cite{kadanoff_1,lasinio_1,lasinio_2,vulpiani_1}, 
which, in contrast,
is invariant 
under any permutation of the spin labels, and 
is expected to have a PDF satisfying
anomalous scaling in the thermodynamic limit 
\cite{lasinio_1,lasinio_2,hilhorst_1,binder_1}.
This suggests to define what we call here a 
``permutation invariant representation'' of the statistics of the 
model. Consider, for instance, the following 
definition of the joint PDF of new exchangeable variables $X_i$'s:
\begin{equation}
p_N(x_1,x_2,\dots,x_N) \equiv 
\frac{1}{N!}\sum_{\pi} \int \prod_{i=1}^N\,d s_i\,
p^\prime_N(s_1,s_2,\dots,s_N)\,\prod_{j=1}^N\,\delta(x_j-s_{\pi(j)}),
\label{eq_ising_ex}
\end{equation}
where the sum is extended to all the $N!$ permutations $\pi$  of
the set $\{1,2,\ldots,N\}$.
The $p_N$'s defined by the projection operation in Eq. (\ref{eq_ising_ex})
are indeed invariant under permutations, while their 
sum $Y_N=\sum_i X_i$ has a PDF identical to that of the
total magnetization $\sum_i S_i$ of the original system. On the basis of 
the same projection, one can also define an effective Boltzmann
factor for the variables $X_i$'s in such a way that the
partition function, and thus the free energy of the original problem, 
are preserved, too. 
Even if the computation of the effective
Hamiltonian in terms of the $X_i$'s is non-trivial,
the above equations show that the asymptotic scaling of the PDF 
of $\sum_i S_i$ for a critical Ising-like model 
and that of $\sum_i X_i$ for its permutation invariant representation, 
coincide. It is also easy to see that one may construct different such
representations of a given statistical model, all sharing the same 
free energy and the same PDF for $Y_N$. 

For a critical Ising system
one expects an anomalous scaling for the PDF of $\sum_{i=1}^N S_i$ with
scaling dimensions $D=15/16$ and $D \simeq 0.825$ for square 
and cubic lattices, respectively \cite{kadanoff_1}. 
Taking into account that finite size scaling for the critical
Ising model implies 
$\langle(\sum_i s_i)^2\rangle_{p^\prime_N} \sim N^{2D}$,
one also concludes that for the permutation invariant representation
defined by Eq. (\ref{eq_ising_ex}) one must have 
$\langle X_i X_j\rangle_{p_N}
\sim N^{2D-2}$ for $N \to \infty$ and $i \neq j$.
As a matter of fact, the limit theorem in
Eq. (\ref{eq_joint_1})
implies a scaling function for the PDF of $Y_N$ and 
linear correlations for the $X_i$'s (Eq. (\ref{eq_lin_corr_1}))
which are compatible
with the asymptotic forms expected for the permutation invariant 
representation of the Ising model constructed here.
 
The above discussion clarifies that correlated variables with
permutation-invariant PDF's can be relevant in the statistical approach 
to anomalous scaling. This relevance stems from the fact that
for these variables the constructive limit theorems presented here are 
valid.
At the same time, additive collective variables like the total 
magnetization of a critical Ising model are considered in many studies 
of complex systems, also outside equilibrium statistical
mechanics \cite{clusel_1}.

\section{Non-Markovian, self-similar stochastic processes}
\label{section_nm_ss_processes}
In many phenomena, anomalous scaling
is a statistical symmetry obeyed to a good approximation
for more or less broad ranges of finite $N$'s.
The validity of limit theorems of the kind proved in the 
previous sections 
opens the possibility of defining joint PDF's
consistent with an exact anomalous scaling of $p_{Y_N}$
for any finite $N$,
i.e. such that $p_{Y_N}=N^D g(N^D y)$.

To illustrate how self-similarity for arbitrary finite $N$ 
arises, let us consider the case of the central limit 
theorem for sums of independent random variables whose 
PDF has finite second moment. The asymptotic scaling is normal 
and turns out to be an attractor in virtue of the 
stability property of the Gaussian PDF.
In particular, this stability implies that if we consider a finite number of
independent increments, $X_1$, $X_2$, ...,$X_N$, each one weighted
by the same Gaussian PDF, the total increment $X_1+X_2+\dots+X_N$
has also precisely a Gaussian PDF, having a width $N^{1/2}$ times
the width of the individual increments. Thus, this PDF
strictly satisfies normal scaling for any $N$.

In an analogous way, the results obtained in the previous
sections for sums of correlated variables allow us to construct joint 
PDF's of the $X$ variables consistent with an exact anomalous scaling
of $p_{Y_N}$, for any finite $N$. 
The generalized stability conditions implied by our limit theorems
make this possible. To be concrete, let us consider the case of the 
scaling function in Eq. (\ref{eq_g_psi}). The construction of Eq. (\ref{eq_joint_psi}) 
implies that if we define
\begin{equation}
p_N(x_1,x_2,\dots,x_N)=
\int_0^{+\infty} d\sigma \int_{-\infty}^{+\infty}
d\mu\;\psi(\sigma,\mu)\,\prod_{i=1}^N\, 
\frac{\exp\left[-\left(x_i/\sigma N^{D-1/2}-\mu/\sigma N^{1/2}\right)^2/2\right]}
{\sqrt{2\pi \sigma^2 N^{2D-1}}},
\label{eq_joint_stable}
\end{equation}
this joint PDF is consistent with an exact anomalous scaling of $p_{Y_N}$
with scaling function $g$ given by Eq. (\ref{eq_g_psi}) and exponent $D>0$,
for any finite $N$.
Since at empirical level $p_{Y_N}$ is often the most accessible PDF of the
system \cite{baldovin_2,stella_1},
such joint PDF's constructed in terms of $g$ may be regarded as a model 
for the dependences determining the anomalous scaling in the range of 
$N$-values relevant for the phenomenon under study.

In the following, let us deal with processes developing in (discrete) time 
and think of $X_i$ as an increment relative to the time 
interval $[(i-1) \Delta t, i \Delta t]$, while the elapsed time
of the process is $t=N \Delta t$ and $\Delta t$ is the elementary 
time-step of the process. Clearly, if $p_N$ is the joint PDF of the first $N$ 
increments of the same process developing in time, causality imposes  
the validity of Eq. (\ref{eq_reduced}) for any $N>1$.  
The conditional PDF 
\begin{equation}
p^c_N(x_N|x_1,x_2,\dots,x_{N-1})\equiv
\frac{p_N(x_1,x_2,\dots,x_N)}
{p_{N-1}(x_1,x_2,\dots,x_{N-1})}
\label{eq_conditional}
\end{equation}
($N\geq2$), expresses the PDF of the $N$-th increment of the process,
conditioned to the history of the previous $N-1$ ones.
Like the joint PDF's, 
the conditional PDF's together with $p_1$ embody the full 
information on the process.
For a causal process with non-Markovian character, 
a property we should be
ready to give up for the $X_i$'s is 
the invariance under permutations of their joint PDF's.

Referring again to an anomalous scaling with $g$
as in Eq. (\ref{eq_g_psi}) and $D>0$, 
it is not difficult to figure out how to modify 
Eq. (\ref{eq_joint_stable}) in order to obtain a discrete-time stochastic process
possessing self-similarity for finite $N$.
To this purpose, let us introduce the following coefficients:
$a_i\equiv[i^{2D}-(i-1)^{2D}]^{1/2}$
and $b_i\equiv i^D-(i-1)^D$,
with $i=1,2,\ldots N,\ldots$. If we then define
\begin{equation}
p_N(x_1,x_2,\dots,x_N)
=\int_0^{+\infty} d\sigma \int_{-\infty}^{+\infty} d\mu\;
\psi(\sigma,\mu)\,\prod_{i=1}^N
\frac{\exp\left[-(x_i-\mu b_i)^2/2\sigma^2 a_i^2\right]}{\sqrt{2\pi \sigma^2 a_i^2}},
\label{eq_joint_stability_time}
\end{equation}
one can verify that this joint PDF indeed
guarantees for any $N$ a strict scaling  
for $p_{Y_N}$:
\begin{equation}
N^D p_{Y_N}(N^D y)=g(y).
\label{eq_scaling_exact}
\end{equation}
Eq. (\ref{eq_scaling_exact}) holds because
the coefficients $a_i$ and $b_i$ satisfy
$\sum_{j=1}^N a_j^2 = N^{2D}$ and $\sum_{j=1}^N b_j =N^D$, respectively.
The condition in Eq. (\ref{eq_reduced}) is also respected.
One recognizes immediately
that for general $\psi(\sigma,\mu)$ the $p_N$'s in Eq. (\ref{eq_joint_stability_time}) are not
permutation invariant anymore for any $D>0$.
The lack of such invariance is also evident in the fact that
$p_{X_i}\equiv p_{X_i,N}$ $\forall N\geq1$ now varies with $i$, reflecting a
nonstationarity of the increments.

When
\begin{equation}
\psi(\sigma,\mu)=\rho(\sigma)\,\delta(\mu)
\label{eq_psi_finance}
\end{equation}
with $\rho(\sigma)\neq\delta(\sigma_0)$, $Y_N$ amounts to a 
stochastic processes of the form
postulated recently for the description of 
financial indexes' evolution \cite{baldovin_2,stella_1}. 
In such a case, the increments are linearly uncorrelated 
and, up to an $i$ dependent rescaling,
their marginal PDF's coincide with $g$.  
The characteristic function of the scaling function $g$
can be expressed as 
$\tilde{g}(k)=\int_0^{\infty} d\sigma \rho(\sigma) 
\exp(-\sigma k^2/2)$, and has the remarkable property that it is converted into 
to a proper $N$-dimensional joint characteristic function if $k$ is replaced
by $\sqrt{k_1^2+k_2^2+\dots+k_N^2}$, for any $N$. Precisely 
this requirement has been identified in Refs. \cite{baldovin_2,stella_1} 
as a natural one for the joint characteristic function 
of the successive returns of an
index. A theorem due to Schoenberg \cite{aldous_1,schoenberg_1} states that the 
$\tilde g(k)$'s having the above form exhaust the class of characteristic 
functions with such property. 
In particular, 
the class of scaling functions from which
one can construct explicit joint PDF's is specified.
This class includes the form
used in Ref. \cite{baldovin_2} and also the Student distribution
recently considered \footnote{
The Authors \cite{challet_1} reach a similar conclusion in a most recent update (v3) 
of their paper.} 
in \cite{challet_1}.

\section{Restoring ergodicity}
\label{section_restoring}
The ergodic properties of the dynamics of stochastic processes like those obtained using
Eqs. (\ref{eq_joint_stability_time},\ref{eq_psi_finance}) 
need to be analyzed in some detail.
To be concrete, let us take $\psi$ as in Eq. (\ref{eq_psi_finance}),
with an arbitrary $\rho$
and $D=1/2$. In this particular case, since the $a_i$'s are
all equal, the increments constitute an exchangeable sequence and are stationary.
Hence, the problem of ergodicity is clearly posed. 
The form of the joint PDF's in 
Eq. (\ref{eq_joint_stability_time}) 
amounts to a convex combination of uncorrelated Gaussian increments
with different $\sigma$'s. 
Any simulation of a single,
infinitely long history $(x_1,x_2,\ldots,x_N,\ldots)$ 
made on the basis of 
the sequence 
$p_1(x_1)= g(x_1),\;p^c_2(x_2|x_1),\;\ldots\;,\;p^c_N(x_N|x_{N-1},\ldots,x_1),\;\ldots$
would not be apt to manifest the ensemble correlations implied by 
$p_N$ in Eq. (\ref{eq_joint_stability_time}). 
Indeed, after an 
initial transient, the extraction of the successive increments 
would essentially be ruled by a Gaussian
conditional PDF with an approximately constant $\sigma=\overline\sigma$, chosen
among all those allowed by $\rho$. 
A different simulation would pick up a different $\overline\sigma$ in the initial
transient stage and then proceed with independent increments
extracted according to this $\overline\sigma$ 
(see Appendix).
The correlations implied by Eq. (\ref{eq_joint_stability_time}) 
are reproduced only by putting together the results of an ensemble of a large
number of different such simulations. A sliding 
time-interval sampling procedure along a single infinite history would not
detect any correlations among the increments.
This amounts to a breaking of ergodicity: The
single infinitely-long realization of the process just isolates one 
of its possible uncorrelated ergodic components, a well 
known consequence of de Finetti's representation theorem for
exchangeable variable sequences \cite{aldous_1}. 
This lack of ergodicity appears at first sight to represent a serious
limitation of the stochastic process, if like in finance a legitimate 
ambition is to simulate single long histories with the same correlation
and scaling properties as the empirical one.

It is possible to recover 
the anomalous scaling and the correlations implied 
by our construction of the joint PDF's 
using a suitably defined dynamics. 
Let us go back to the motivations mentioned
above for considering self-similar processes: The
approximate satisfaction of anomalous scaling 
for PDF's like that of the aggregated increment in a time interval of
duration $\tau$ is often valid for a limited range, 
$\tau\leq M\,\Delta t$.
Under these premises, an adequate goal for the simulation is that of
reproducing, by time-averages along a single dynamical trajectory, 
the scaling and correlation properties
implied by Eq. (\ref{eq_joint_stability_time})
just over the time range $M\,\Delta t$. 
One way of obtaining these properties, 
namely ergodicity
and self-similarity up to the
time-scale $M\,\Delta t$,  
is by implementing an autoregressive dynamics \cite{engle_1} with 
memory span equal to $M$.
Imagine we have extracted, consistently with the conditional
PDF's $p^c_i$, $i=1,2,\ldots,M$, the first
$M$ increments of the additive variable $Y_M$. Instead of using
the conditional PDF $p^c_{M+1}(x_{M+1}|x_M,x_{M-1},\ldots,x_1)$ 
to extract the $M+1$-th increment, 
we use $p^c_M(x_{M+1}|x_M,x_{M-1},\ldots,x_2)$.
Similarly, for any time $t>M\,\Delta t$ we use this autoregressive
scheme in which only the
preceding $M-1$ increments have an effect on the further evolution.
In this way one circumvents the problem of broken
ergodicity, because for finite $M$ the conditioning input
is constantly updated and modified to an extent which is sufficient 
for a long-enough simulation to span all the $\sigma$'s allowed by the 
ensemble in Eq. (\ref{eq_joint_stability_time}). 
With such strategy
the empirical PDF of the sum of the increments over an interval $\tau$,
sampled from all intervals of duration $\tau$ along a single long history of the process,
satisfies to a very good
approximation the anomalous scaling for $\tau\leq M\,\Delta t$  
(see Appendix).

\section{Concluding remarks and perspectives}
In this Article we have shown that the choice of 
variables with joint PDF's invariant under permutations is 
particularly favorable for discussing the
problem of the asymptotic emergence and universality of 
anomalous scaling due to correlations. 
Ideas of the modern theory of critical phenomena and complex
systems are at the basis of 
the advancements we could present here. 
Our limit theorems cover indeed forms of anomalous
scaling, which, to our best knowledge, so far have not been treated by the
probabilistic literature with the present generality.  
At the same time, classical examples taken from the theory of critical 
phenomena gave us a way to illustrate the role 
variables with permutation invariant joint PDF's can play 
in more general problems with anomalous scaling.

As remarked above, the idea of basing limit theorems for
correlated variables on some suitable generalization of
the standard multiplication has some appeal
\cite{baldovin_1}. 
The rules by which we compose the $l$ PDF's to
obtain $p_N$ in
Eqs. (\ref{eq_joint_1}) or (\ref{eq_joint_psi}), 
retain in fact the commutative and
associative properties. 
In this respect, our approach to anomalous scaling is quite different
from the renormalization group one, and remains closer in spirit
to the limit theorems for independent variables. 
This closeness is also manifest in the relative simplicity
of our proofs, which directly rely on the corresponding ones for the
independent case. 
Thus, the mathematics at the basis of the standard
central limit theorem plays a fundamental role 
also outside the context of independent variables. 
Another difference of our approach compared to the renormalization
group is that we do not need to make use of the hierarchical
modeling to have analytical control on 
statistical coarse-graining operations. Here we replace the
hierarchical paradigm by the assumption of invariance under
permutations.
In principle, this replacement still allows to address realistic
scalings as illustrated in Section \ref{section_permutation}. 

We have expressed our limit PDF's for the (rescaled) sums of 
correlated random variables  as convex combinations of 
Gaussians with varying widths and/or centers. Scaling functions 
belonging to this class have been considered very often in 
phenomenological descriptions of anomalous scaling \cite{gheorghiu_1}, 
but their possible implications as far as correlations are concerned 
were not stressed enough, in our opinion.
The wide classes of scaling functions and the continuous ranges of 
scaling exponents identified through our theorems, definitely do not
support the idea that in the context of strongly correlated
variables relevant scaling forms could be organized in a restricted
set of universality classes. 
In particular, there does not appear to exist one or few particular
scaling functions playing a universal role similar to the one of the
Gaussian in the independent case. 

The generalization of the notion of stability implied by our theorems
naturally leads to the introduction of self-similar stochastic
processes with correlated increments. These include in particular
the process proposed in Ref. \cite{baldovin_2} as a model of index evolution
in finance. Besides giving this proposal a rigorous basis, the results 
presented here, especially those concerning the restoration of
ergodicity, substantially enhance the analytical and numerical 
tractability of such a process.

\begin{acknowledgments}
We acknowledge Giovanni Jona-Lasinio for useful discussions 
and comments on the manuscript.
This work is supported by 
``Fondazione Cassa di Risparmio di Padova e Rovigo'' within the 
2008-2009 ``Progetti di Eccellenza'' program. 
\end{acknowledgments}

\section{Appendix}

In the first part of this Appendix, we prove
three different statements which in particular imply that $p_{Y_N}$, the  
PDF of 
$Y_N\equiv\sum_{i=1}^N X_i$ satisfies the scaling 
\begin{equation}
N^D\:p_{Y_N}(N^Dy)\to g(y), 
\label{eq_scaling_ap}
\end{equation}
for $N\to\infty$ (refer to main text for details).

\bigskip
\noindent{\bf Limit Theorem for $g$'s given by Gaussian mixtures with different 
centers and $D=1/2$}\\
{\it 
Given the sequence 
of joint PDF's 
\begin{equation}
p_N(x_1,x_2,\dots,x_N) = \int_{-\infty}^{+\infty} d\mu\;
\lambda(\mu)\;\prod_{i=1}^N\;
l\left(x_i-\frac{\mu}{N^{1-D}}\right),\quad N=1,2,\ldots
\label{eq_g_lambda_ap}
\end{equation}
for the random variables $\left\{X_i\right\}_{i=1,2,\ldots,N}$,
where $D=1/2$, $\lambda$ and $l$ are single-variable PDF's 
with $\langle\mu\rangle_\lambda=0$ and 
$\langle X\rangle_l=0$, $\langle X^2\rangle_l=1$,
then as $N\to\infty$ the probability 
\begin{equation}
Prob\left\{\sum_{i=1}^N \frac{X_i}{N^{D}} <z\right\}\to
\int_{-\infty}^z d w\;g(w)
\end{equation}
uniformly, 
with 
\begin{equation}
g(w)=\int_{-\infty}^{+\infty} d\mu\;\lambda(\mu)\;
\frac{\exp\left[-(w-\mu)^2/2\right]}{\sqrt{2\pi}}.
\end{equation}
}

\smallskip
Let us consider the positive quantity
\begin{equation}
Prob\left\{\sum_{i=1}^N \frac{X_i}{N^{1/2}} <z\right\}_{\mu} \equiv
\int_{-\infty}^z d w \;\int \prod_{i=1}^Nd x_i\;
l\left(x_i-\frac{\mu}{N^{1/2}}\right)\;
\delta\left(w-\sum_{i=1}^N \frac{x_i}{N^{1/2}}\right),
\label{eq_prob_1}
\end{equation}
which, once multiplied by $\lambda$ and integrated with respect to $\mu$, yields 
the probability that $\sum_i X_i/N^{1/2} \leq z$. 
The following identity holds:
\begin{equation}
Prob\left\{\sum_{i=1}^N \frac{X_i}{N^{1/2}} <z\right\}_{\mu} =
Prob\left\{\sum_{i=1}^N \frac{X_i}{N^{1/2}} < z -\mu \right\}_{0}.
\label{eq_prob_2}
\end{equation}
The central limit theorem for independent variables guarantees 
\cite{gnedenko_1,feller_1}
that the right hand side of Eq. (\ref{eq_prob_2}) converges uniformly 
to
\begin{equation}
\int_{-\infty}^z d w \frac{\exp\left[-(w-\mu)^2/2\right]}{\sqrt{2\pi}}.
\end{equation}
Since the uniform convergence holds for $z$ and $\mu$ separately,
we can interchange the integration in $\mu$ with the limit for $N \to \infty$
and get 
\begin{equation}
\int_{-\infty}^{+\infty} d\mu \lambda(\mu) Prob\left\{\sum_{i=1}^N 
\frac{X_i}{N^{1/2}} <z \right\}_{\mu} \to 
\int_{-\infty}^z d w \int_{-\infty}^{+\infty} d\mu \lambda(\mu) 
\frac{\exp\left[-(w-\mu)^2/2\right]}{\sqrt{2\pi}},
\label{eq_limit_mu}
\end{equation}
still uniformly in $z$.
This proves the asymptotic scaling (\ref{eq_scaling_ap}) of $p_{Y_N}$,
with $D=1/2$ and $g$ as in Eq. (\ref{eq_g_lambda_ap}).

\bigskip
\noindent{\bf Limit Theorem for $g$'s given by Gaussian mixtures with different 
centers and $D>1/2$}\\
Here, we establish a similar result with $D>1/2$. 
Let us
look back at Eq. (\ref{eq_g_lambda_ap}), where we have a convex combination of
Gaussians with finite second moment equal to $1$. Suppose to perform a limit
in which this second moment is sent to zero: In this limit the Gaussian 
would approach a Dirac delta-function. Hence, we would have 
\begin{equation}
g(x)=\lambda(x).
\end{equation}
In order to construct $p_N$  such that $p_{Y_N}$
satisfies asymptotic scaling with
$g=\lambda$ and with $D>1/2$, it is convenient to
consider the characteristic functions of $p_N$, and of $g$, respectively:
\begin{eqnarray}
\tilde{p}_N(k_1,\dots,k_N)&=& \int \prod_{i=1}^N\,d x_i\,\exp{(-i k_i x_i)}\;p_N(x_1,\dots,x_N),\\
\tilde{g}(k) &=& \int d w\,\exp{(-i k w)}\,g(w).
\end{eqnarray}
We can prove the following statement.\\
{\it 
Given the sequence 
of joint PDF's 
\begin{equation}
p_N(x_1,x_2,\dots,x_N) = \int_{-\infty}^{+\infty} d\mu\;
\lambda(\mu)\;\prod_{i=1}^N\;
l\left(x_i-\frac{\mu}{N^{1-D}}\right),\quad N=1,2,\ldots
\end{equation}
for the random variables $\left\{X_i\right\}_{i=1,2,\ldots,N}$,
where $D>1/2$, $\lambda$ and $l$ are single-variable PDF's 
with $\langle\mu\rangle_\lambda=0$ and 
$\langle X\rangle_l=0$, $\langle X^2\rangle_l=1$,
then as $N\to\infty$ we have  
\begin{equation}
\tilde{p}_N\left(\frac{k}{N^D},\frac{k}{N^D},\dots,\frac{k}{N^D}\right) 
\to \tilde{g}(k),
\label{eq_scaling_cf}
\end{equation}
with 
\begin{equation}
g(w)=\lambda(w).
\end{equation}
The convergence 
is uniform in $k$ if $|\tilde\lambda(k)|$ decays at large $|k|$ 
as $1/|k|^2$ or faster, uniform in $k$ in any bounded subset of $\mathbb R$ otherwise.
}

\smallskip
Indeed, the characteristic function of such $p_N$ is
\begin{equation}
\tilde{p}_N(k_1,\dots,k_N)=
\int_{-\infty}^{+\infty} d\mu\,\lambda(\mu)\,\exp\left[-i(k_1+\dots+k_N)
  \frac{\mu}{N^{(1-D)}}\right]\,\prod_{i=1}^N\,\tilde{l}(k_i),
\label{eq_joint_cf_D}
\end{equation}
where $\tilde{l}$ is the characteristic function of $l$. We can write
\begin{equation}
\tilde{p}_N\left(\frac{k}{N^D},\frac{k}{N^D},\dots,\frac{k}{N^D}\right)=
\int_{-\infty}^{+\infty} d\mu\,\lambda(\mu)\,\exp{(-i k \mu)}\;\tilde{l}(k/N^D)^N.
\end{equation}
If we assume $D > 1/2$, $\tilde{l}(k/N^D)^N$ approaches $1$ for $N \to
\infty$, uniformly in $k$ in any bounded subset of $\mathbb R$. 
This implies that 
as $N\to\infty$,
\begin{equation}
\tilde{p}_N\left(\frac{k}{N^D},\frac{k}{N^D},\dots,\frac{k}{N^D}\right)\to
\tilde\lambda(k),
\label{eq_limit_lambda}
\end{equation}
which proves the theorem.
The convergence in Eq. (\ref{eq_limit_lambda}) 
is uniform in $k$ if $|\tilde\lambda(k)|$ decays at large $|k|$ 
as $1/|k|^2$ or faster, uniform in $k$ in any bounded subset of $\mathbb R$ otherwise.

\bigskip
\noindent{\bf Limit Theorem for $g$'s given by Gaussian mixtures 
with different centers and widths, and $D>0$}\\
{\it 
Given the sequence 
of joint PDF's 
\begin{equation}
p_N(x_1,\dots,x_N)=
\int_0^{\infty} d\sigma \int_{-\infty}^{+\infty} d\mu\,
\psi(\sigma,\mu)\,\prod_{i=1}^N\,
\frac{l(x_i/\sigma N^{D-1/2}-\mu/\sigma N^{1/2})}{\sigma N^{D-1/2}},
\label{eq_joint_psi_ap}
\end{equation}
for the random variables $\left\{X_i\right\}_{i=1,2,\ldots,N}$,
where 
$D>0$, 
$\psi$ is a joint PDF identically equal to zero in a whole neighborhood of
$\sigma=0$ and such that $\langle\mu\rangle_\psi=0$,
and $l$ is a single-variable PDF 
with 
$\langle X\rangle_l=0$, $\langle X^2\rangle_l=1$,
then as $N\to\infty$ the probability 
\begin{equation}
Prob\left\{\sum_{i=1}^N \frac{X_i}{N^{D}} <z\right\}\to
\int_{-\infty}^z d w\;g(w)
\end{equation}
uniformly, 
with 
\begin{equation}
g(w)= \int_0^{+\infty} d\sigma \int_{-\infty}^{+\infty} d\mu\,\psi(\sigma,\mu)\,
\frac{\exp\left[-(w-\mu)^2/{2\sigma^2}\right]}{\sqrt{2\pi \sigma^2}}.
\label{eq_g_psi_ap}
\end{equation}
}

\smallskip
For any $\sigma$ and $\mu$ we define the quantity
\begin{eqnarray}
Prob\left\{\sum_{i=1}^N X_i/N^{D}<z\right\}_{\sigma,\mu}&\equiv&
\int_{-\infty}^z d w 
\int_{-\infty}^{+\infty} d x_1 \dots d x_N
\delta\left(w-\frac{x_1+\dots +x_N}{N^{D}}\right)
\prod_{i=1}^N \frac{l(x_i/\sigma N^{D-1/2}-\mu/\sigma N^{1/2})}{\sigma N^{D-1/2}}\nonumber\\
&\equiv&f\left(z,N,D,\mu,\sigma\right)
\label{eq_given_sigma}
\end{eqnarray}
One easily verifies the following property:
\begin{equation}
f\left(z,N,D,\mu,\sigma\right)=
f\left(\frac{z}{\sigma}-\frac{\mu}{\sigma},N,1/2,0,1\right).
\label{eq_sigma_1}
\end{equation}
Under the present assumptions, the central limit theorem for independent variables 
\cite{gnedenko_1,feller_1}  guarantees that
the quantity on the right hand side of Eq. (\ref{eq_sigma_1}) converges
uniformly to the limit 
\begin{equation}
\int_{-\infty}^z d w \frac{\exp(-(w-\mu)^2/2\sigma^2)}{\sqrt{2\pi \sigma^2}}.
\label{eq_uniformly}
\end{equation}
In view of the conditions on $\psi$, this uniformity holds
for $\sigma$, $\mu$ and $z$, separately.
We thus conclude that
\begin{equation}
\int_0^\infty d\sigma \int_{-\infty}^{+\infty} d\mu\,\psi(\sigma,\mu)\,
 Prob\left\{\sum_{i=1}^N X_i/N^{1/2} < z\right\}_{\sigma,\mu} \to
\int_{-\infty}^z d w \int_0^{\infty} d\sigma \int_{-\infty}^{+\infty} d\mu\,
\psi(\sigma,\mu)\,\frac{\exp(-(w-\mu)^2/2\sigma^2)}
{\sqrt{2 \pi \sigma^2}},
\label{eq_theorem}
\end{equation}
uniformly in $z$. 
The uniformity follows from the 
hypothesis that $\psi$ is zero in a whole  neighborhood of $\sigma=0$. 

If $\psi(\sigma,\mu)=\rho(\sigma) \delta(\mu)$,
i.e. with $g$ given by a mixture of Gaussians of different 
widths and all centered in the origin, the $X$ 
variables are linearly uncorrelated for any $N$.

\bigskip

\noindent{\bf Simulating a self-similar process with strongly correlated increments}\\
Here we discuss the issue of how to simulate the
stochastic processes described in the main text.
Let $X_i$ be the increment relative to the time interval $[(i-1)\Delta t,i\Delta t]$
of the discrete-time process $Y_N\equiv \sum_{i=1}^N X_i$, where
$\Delta t$ is an elementary time-step and $t\equiv N\,\Delta t$ the
elapsed time. 
For the sake of definiteness, we assume $\psi$ to be of the form
\begin{equation}
\psi(\sigma,\mu)=\delta(\mu)\;\rho(\sigma),
\label{eq_psi}
\end{equation}
with 
\begin{equation}
\rho(\sigma)=A\frac{6 b^3\sigma^2}{\pi\left(b^6+\sigma^6\right)}
\label{eq_rho}
\end{equation} 
for $\sigma\in(\sigma_{min},\sigma_{max})$ 
($0<\sigma_{min}<\sigma_{max}$) and $\rho(\sigma)=0$ elsewhere. 
The parameter $A$ is just a normalization constant fixed such as 
$\int_{\sigma_{min}}^{\sigma_{max}}d\sigma\,\rho(\sigma)=1$, 
whereas $b$ determines $\langle\sigma^2\rangle_\rho$. 
With such a choice, the scaling function 
\begin{equation}
g(x)= \int_{\sigma_{min}}^{\sigma_{max}}d\sigma\;
\frac{6 b^3\sigma^2}{\pi\left(b^6+\sigma^6\right)}\;
\frac{\exp\left[-x^2/{2\sigma^2}\right]}{\sqrt{2\pi \sigma^2}},
\label{eq_g_rho}
\end{equation}
is even. 
With a
sufficiently large $\sigma_{max}$, we can mimic a fat-tail
power-law decay for $g$ of the kind $g(x)\sim1/|x|^4$ at large arguments.
We first address the situation in which the problem 
of breaking of ergodicity is well posed, i.e., when the increments
$X_i$'s of the process are stationary, so that it makes sense to 
compare their ensemble and time averages. 
Later, we will comment about the more general case. 
We thus fix $D=1/2$. 
With the choice (\ref{eq_psi}) for $\psi$, this 
also implies that the $X_i$'s are exchangeable. 
Indeed, according to Eq. (\ref{eq_joint_stability_time}) of the main text, 
the joint PDF for the increments of the process becomes 
\begin{equation}
p_N(x_1,\ldots,x_N)=\int_{\sigma_{min}}^{\sigma_{max}} d\sigma \frac{6 b^3\sigma^2}{\pi\left(b^6+\sigma^6\right)}
\frac{e^{-(x_1^2+\cdots+x_N^2)/2\sigma^2}}
{\left(2\pi\sigma^2\right)^{N/2}},
\end{equation}
and a straightforward calculation yields
\begin{eqnarray}
C_{\alpha\beta}(i,j)&\equiv&
\frac{
\langle|X_i|^\alpha|X_j|^\beta\rangle_{p_N}-
\langle|X_i|^\alpha\rangle_{p_1}\langle|X_j|^\beta\rangle_{p_1}
}{
\langle|X_i|^{\alpha+\beta}\rangle_{p_1}-
\langle|X_i|^\alpha\rangle_{p_1}\langle|X_i|^\beta\rangle_{p_1}
}\\
&=&
\frac{
B_{\alpha}B_{\beta}
\;
\left[
\langle\sigma^{\alpha+\beta}\rangle_{\rho} - 
\langle\sigma^{\alpha}\rangle_{\rho}
\langle\sigma^{\beta}\rangle_{\rho}
\right]
}{
B_{\alpha+\beta}
\langle\sigma^{\alpha+\beta}\rangle_{\rho} -
B_{\alpha}B_{\beta}
\;
\langle\sigma^{\alpha}\rangle_{\rho}
\langle\sigma^{\beta}\rangle_{\rho}
}
\label{eq_correlator}
\end{eqnarray}
$\forall\;i,j=1,2,\ldots,N$, with 
\begin{equation}
  B_{\alpha}\equiv\int_{-\infty}^{+\infty}d x\;|x|^\alpha\; 
\frac{e^{-x^2/2}}
{\sqrt{2\pi}}.
\end{equation}
Notice that when  
$\sigma_{max}\to\infty$, $\langle\sigma^{\alpha+\beta}\rangle_{\rho}$
is finite only for $\alpha+\beta<3$.
The strong correlations among the increments are reflected
by the fact that the $C_{\alpha\beta}(i,j)$ is different
from zero. 
On the other hand, $\langle X_i X_j\rangle_{p_N}=0$
$\forall\;j\neq i$, and the process is uncorrelated at linear level. 
Hence, 
\begin{equation}
C_{l i n}(i,j)\equiv
\frac{
\langle X_i X_j\rangle_{p_N}-
\langle X_i\rangle_{p_1}\langle X_j\rangle_{p_1}
}{
\langle X_i^2\rangle_{p_1}-
\langle X_i\rangle_{p_1}\langle X_i\rangle_{p_1}
}
\end{equation}
is equal to $1$ for $j=i$, and zero otherwise. 
 
\begin{figure} 
\includegraphics[width=0.49\columnwidth]{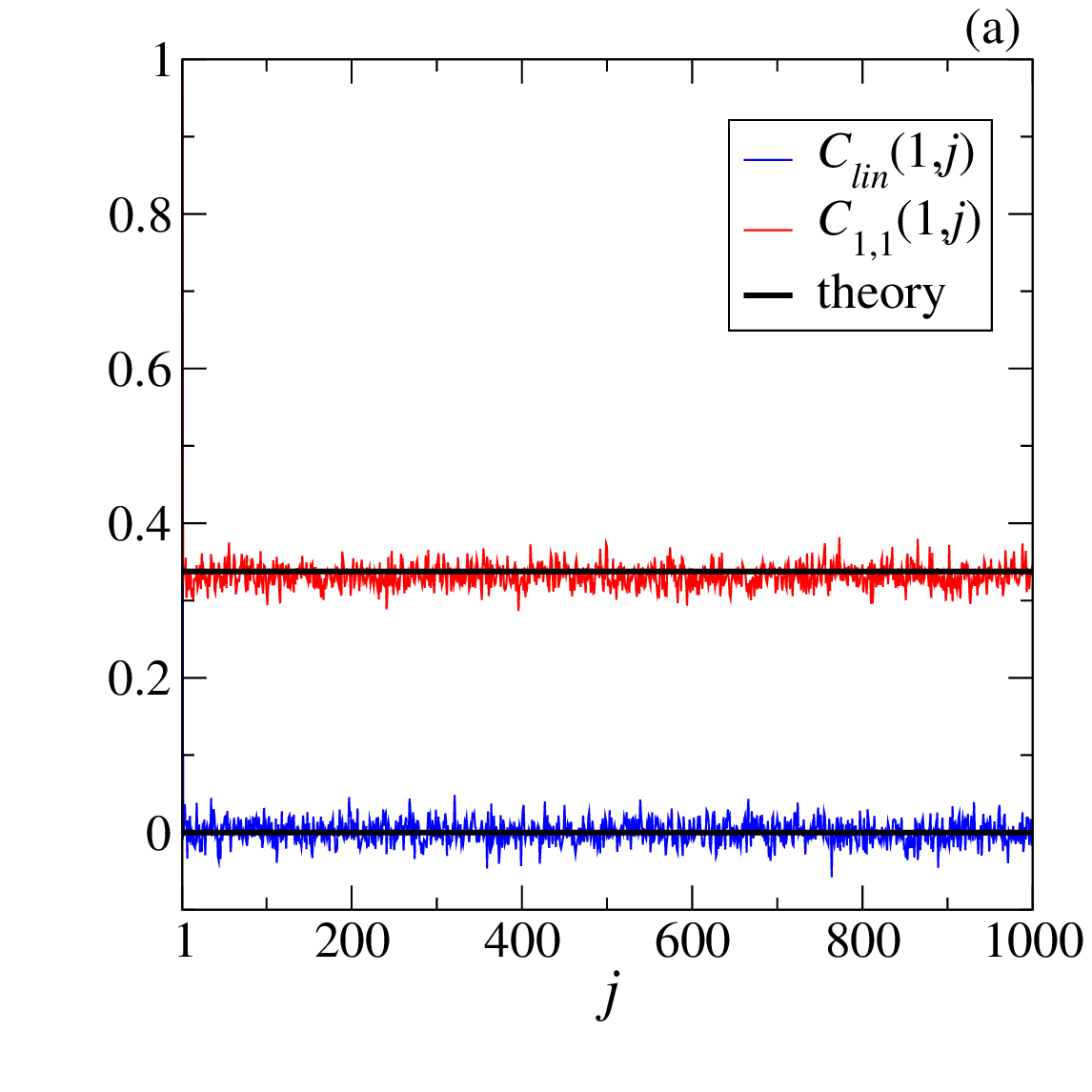}
\includegraphics[width=0.49\columnwidth]{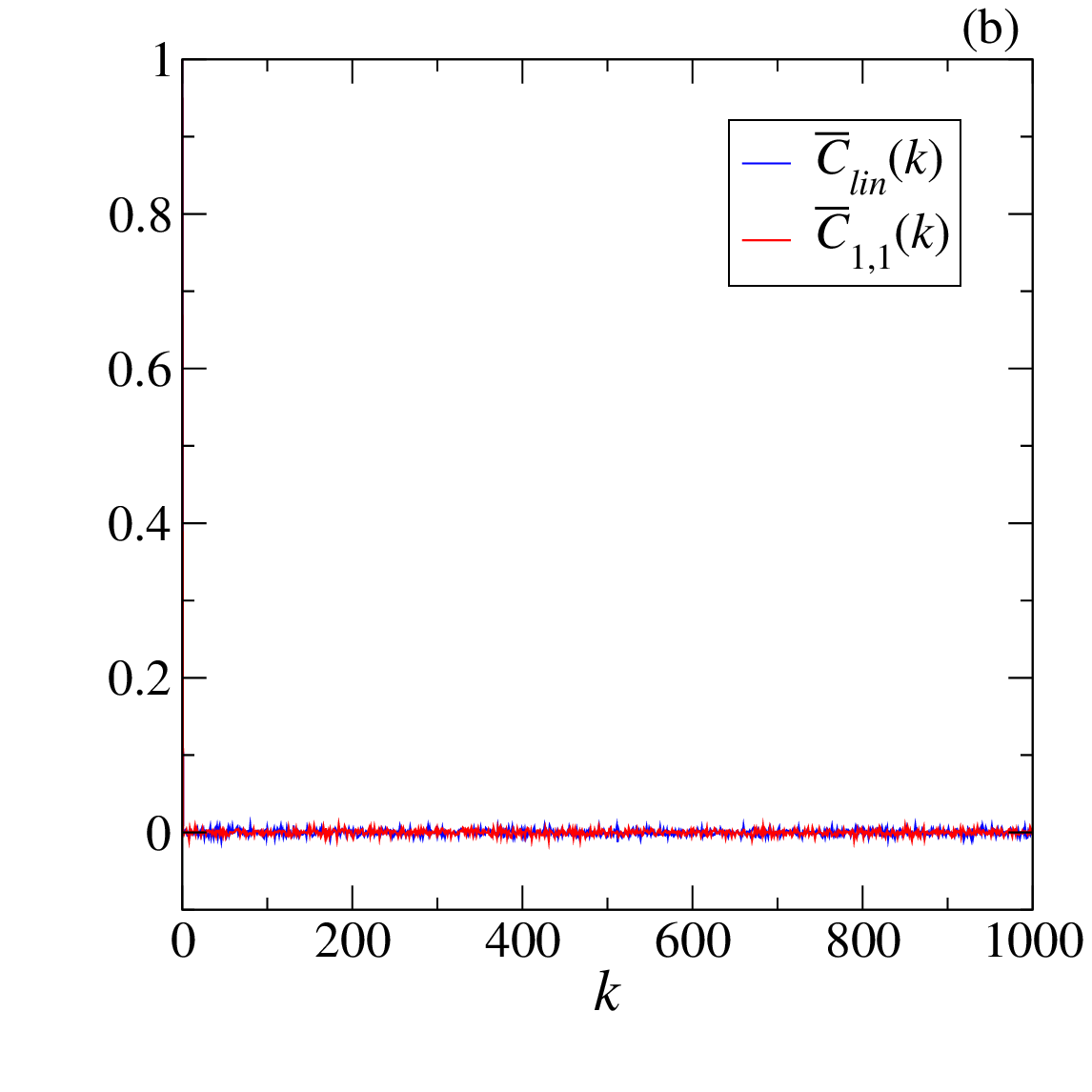}
\caption{
  Ergodicity breaking for a progressive simulation. 
  Correlations calculated from an ensemble of $10^5$ realizations (a) and 
  from a single realization of $10^5$ steps (b). Here, and in the
  following we use $\rho(\sigma)$
  as in Eq. (\ref{eq_rho}), with $\sigma_{min}=0.01$,
  $\sigma_{max}=10$, and $b=1/\sqrt{2}$. 
}
\label{fig_prog_corr}
\end{figure}

A natural strategy of simulation of the process is based on extracting
the random increments $x_1,x_2,\ldots,x_N,\ldots$ according to
the sequence of conditional PDF's
\begin{equation}
p_1(x_1)=g(x_1),\;p_2^c(x_2|x_1),\;\ldots,\;p_i^c(x_i|x_{i-1},\ldots,x_1),\;\ldots,
\label{eq_progressive}
\end{equation}
respectively. 
To distinguish with what follows, we call this simulation scheme ``progressive''. 
An ensemble of a large number of independent simulations of
this kind reproduces well all the theoretical features of the
process. For instance, in Fig. \ref{fig_prog_corr}a we find that $C_{l i n}(1,j)$ 
and $C_{1,1}(1,j)$  
obtained from an ensemble of $10^5$
simulations oscillate around the correct theoretical values.
On the contrary, if we consider a single simulation of $N$ steps
($N\gg1$) generated
according to Eq. (\ref{eq_progressive}), and the associated ``sliding-window'' 
correlators 
\begin{eqnarray}
\overline{C}_{\alpha\beta}(k)&\equiv&
\frac{
\frac{1}{N-k}\sum_{i=1}^{N-k}
|x_i|^\alpha|x_{i+k}|^\beta-
\left(\frac{1}{N}\sum_{i=1}^{N}
|x_i|^\alpha\right)
\left(\frac{1}{N-k}\sum_{i=1}^{N-k}
|x_{i+k}|^\beta\right)
}{
\frac{1}{N}\sum_{i=1}^{N}
|x_i|^{\alpha+\beta}-
\left(\frac{1}{N}\sum_{i=1}^{N}
|x_i|^\alpha\right)
\left(\frac{1}{N}\sum_{i=1}^{N}
|x_i|^\beta\right)
},\\
\overline{C}_{l i n}(k)&\equiv&
\frac{
\frac{1}{N-k}\sum_{i=1}^{N-k}
x_i x_{i+k}-
\left(\frac{1}{N}\sum_{i=1}^{N}
x_i\right)
\left(\frac{1}{N-k}\sum_{i=1}^{N-k}
x_{i+k}\right)
}{
\frac{1}{N}\sum_{i=1}^{N}
x_i^2-
\left(\frac{1}{N}\sum_{i=1}^{N}
x_i\right)
\left(\frac{1}{N}\sum_{i=1}^{N}
x_i\right)
},
\end{eqnarray}
we find that both $\overline{C}_{1,1}(k)$ and 
$\overline{C}_{l i n}(k)$ are zero for $k>0$ (see Fig. \ref{fig_prog_corr}b). 
This means that time-averages disagree with ensemble-averages, i.e., the
dynamics is not ergodic. 

We gain an insight into this ergodicity breaking by noticing that
the conditional PDF for the next increment at
each time-step $i$ can be expressed in the following way:
\begin{equation}
p_i^c(x_i|x_{i-1},\ldots,x_1)=\int_{\sigma_{min}}^{\sigma_{max}}d\sigma\;
\rho_i^c(\sigma|x_{i-1},\ldots,x_1)\;
\frac{e^{-x_i^2/2\sigma^2}}{\sqrt{2\pi\sigma^2}},
\label{eq_conditional_ap}
\end{equation}
where the conditional PDF for the value $\sigma$, $\rho_i^c$, is in fact a
function $f_i$ depending only on $\sum_{j=1}^{i-1} x_j^2$:
\begin{equation}
\rho_i^c(\sigma|x_{i-1},\ldots,x_1)=
\frac{
\frac{\rho(\sigma)\;\prod_{j=1}^{i-1}e^{-x_j^2/2\sigma^2}}
{\sigma^{M}}
}
{
\int_0^{+\infty}d\sigma^\prime\;
\frac{\rho(\sigma^\prime)\;\prod_{j=1}^{i-1}e^{-x_j^2/2\sigma^{\prime2}}}
{\sigma^{\prime M}}
}
\equiv f_i\left(\sigma,\sum_{j=1}^{i-1} x_j^2\right).
\label{eq_sigma_cond_ap}
\end{equation}
As $i$ increases, very quickly $f_i$ becomes sharply peaked 
around a specific value $\overline\sigma$, which depends on the sum of the
squares of the past increments, 
$\sum_{j=1}^{i-1} x_j^2$.
For a given $i\gg1$, the dynamics is such that the typical growth of 
$\sum_{j=1}^{i} x_j^2$ with respect to $\sum_{j=1}^{i-1} x_j^2$ compensates 
for the functional change of $f_{i+1}$ with respect to $f_{i}$,
and the new
conditional PDF, $\rho_{i+1}^c$, remains peaked around the same
value $\overline\sigma$. 
In this way, a single ergodic component labeled by $\overline\sigma$
is chosen during the initial stages of the simulation, when 
$\rho_{i}^c$ still resembles $\rho$. 
The subsequent dynamical evolution is then very similar to a process with
independent increments at the initially selected $\overline\sigma$. 
This is why along a single history of the
process the sliding-window analysis performed in Fig. \ref{fig_prog_corr}b reveals
a vanishing $\overline C_{1,1}(k)$ for $k>0$.

\begin{figure} 
\includegraphics[width=0.49\columnwidth]{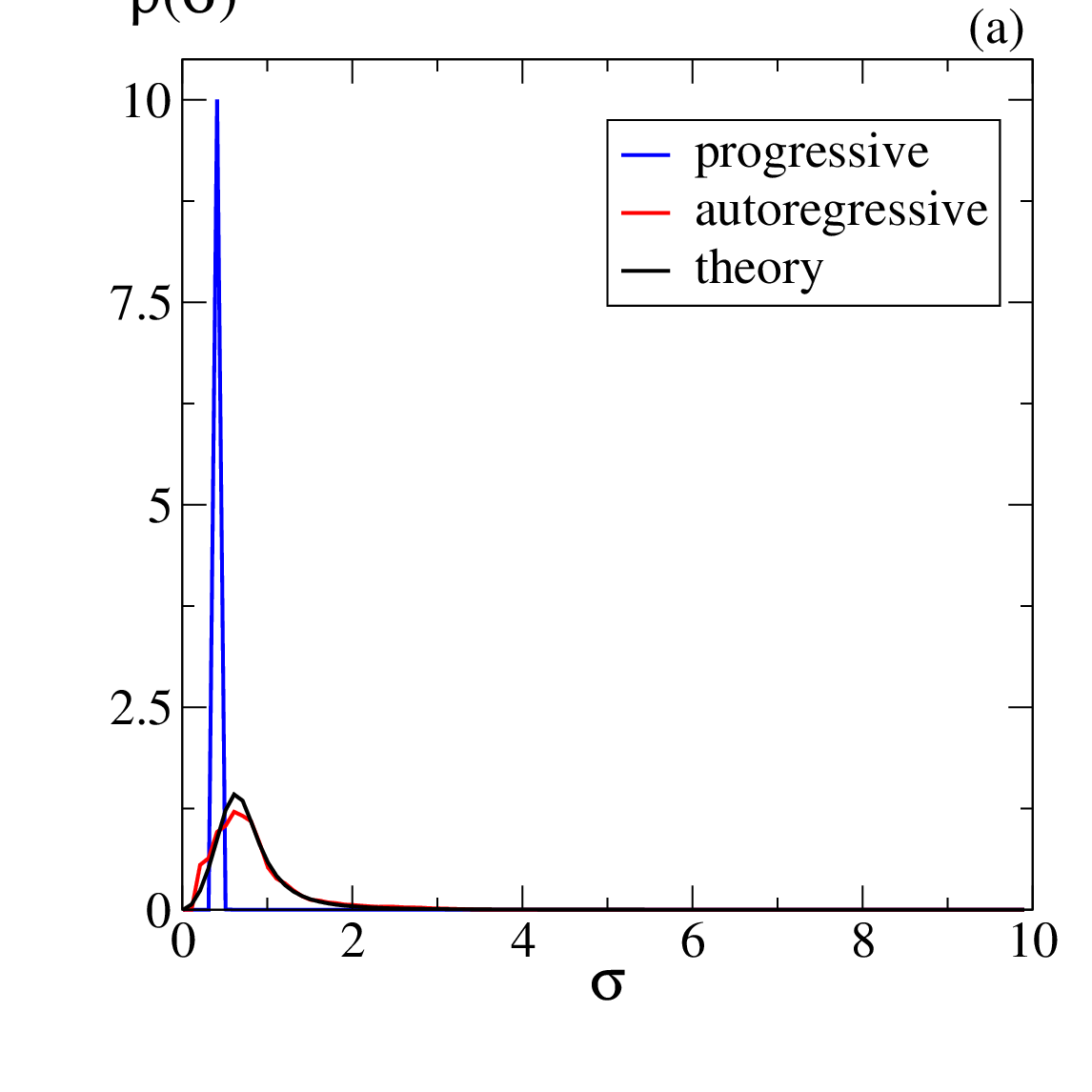}
\includegraphics[width=0.49\columnwidth]{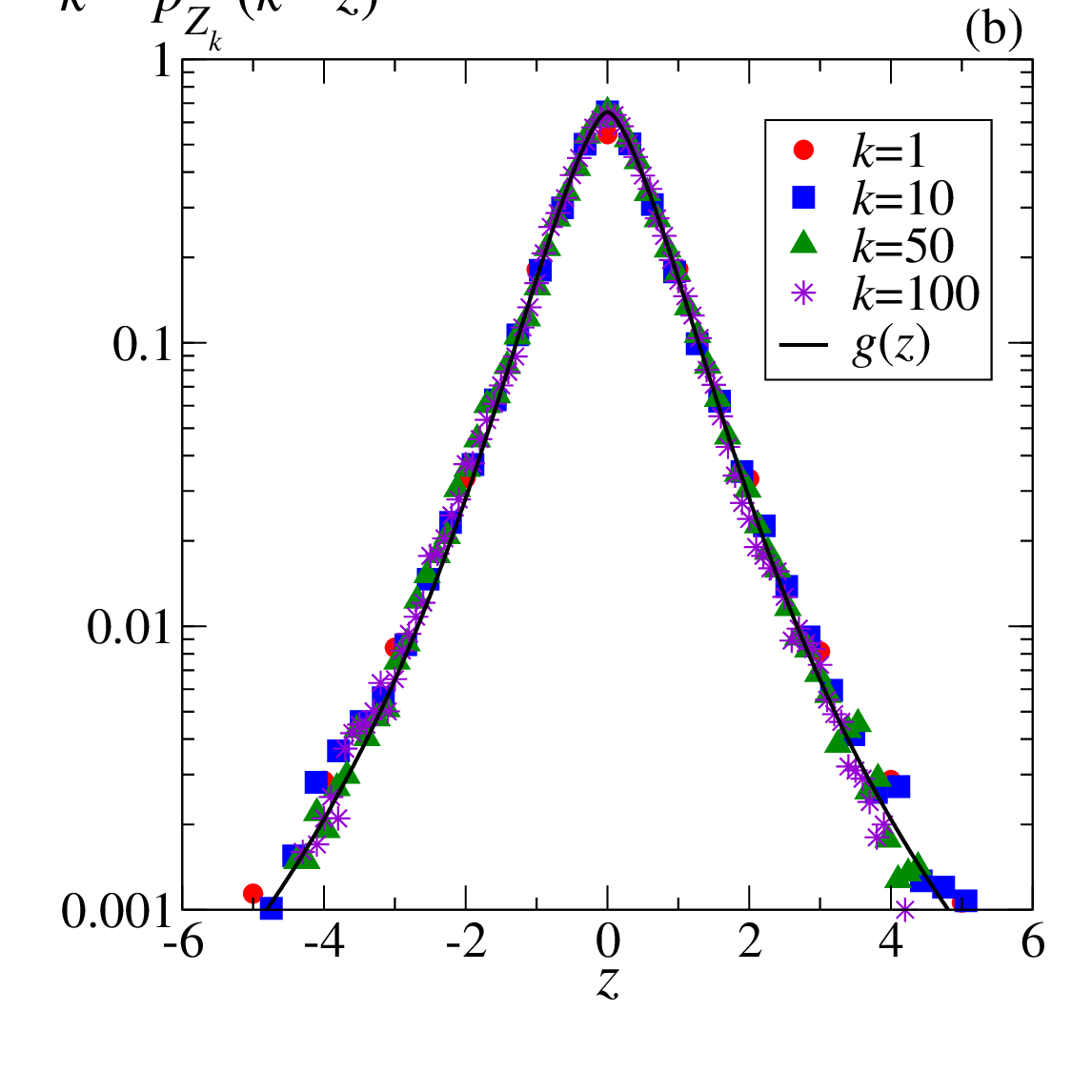}
\caption{
  (a) Histogram of the frequency of ergodic components $\sigma$ in $\rho_i^c$:
  Only for the autoregressive simulation the histogram reproduces well
  $\rho(\sigma)$ in Eq. (\ref{eq_rho}). 
  (b) The rescaling of the histogram of the 
  increments over an interval of duration $\tau=k\,\Delta t$ for a
  single autoregressive simulation of $10^5$ time-steps
  with $M=100$ reproduces $g(z)$ for $k\leq M$.
}
\label{fig_histo_scaling}
\end{figure}

\begin{figure} 
\includegraphics[width=0.49\columnwidth]{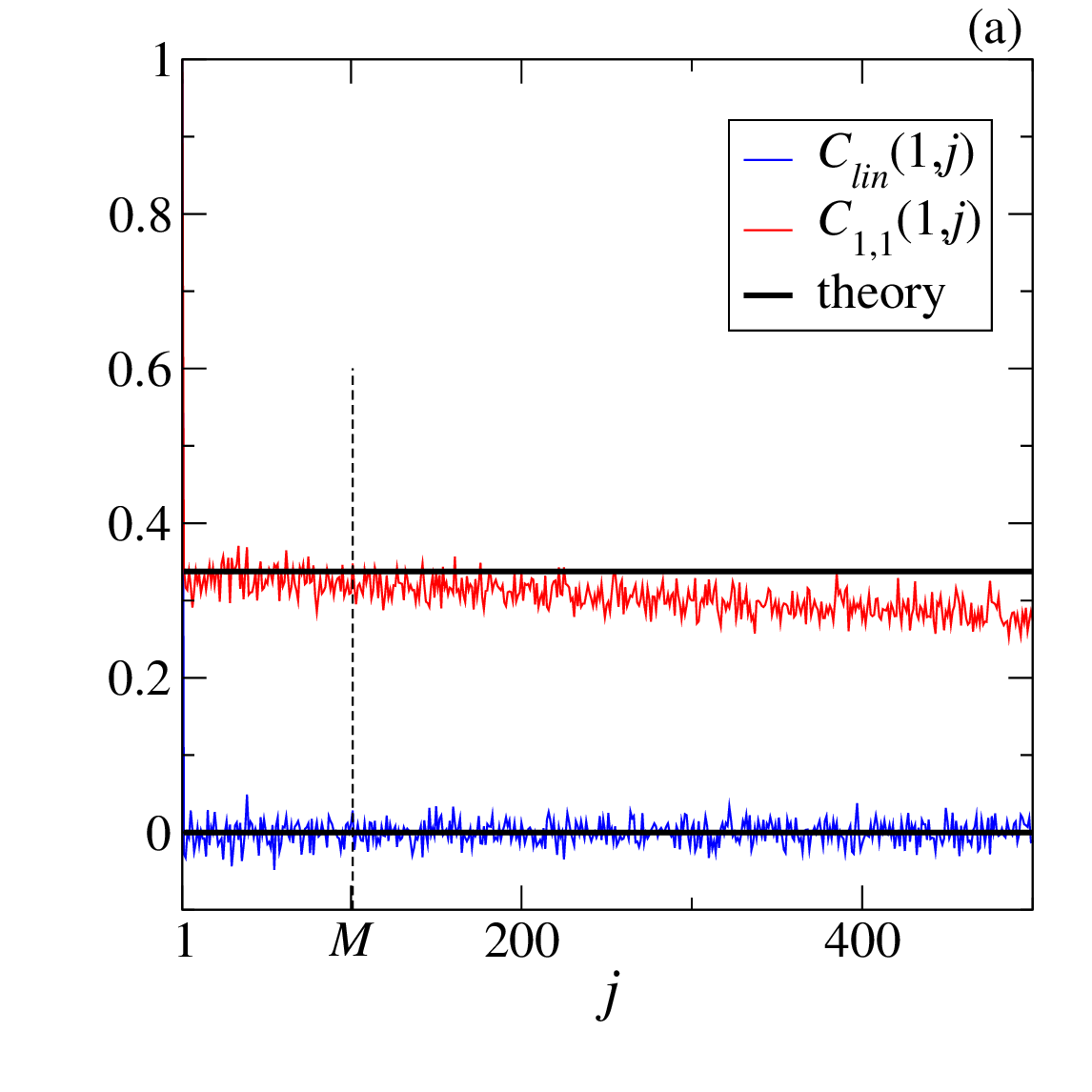}
\includegraphics[width=0.49\columnwidth]{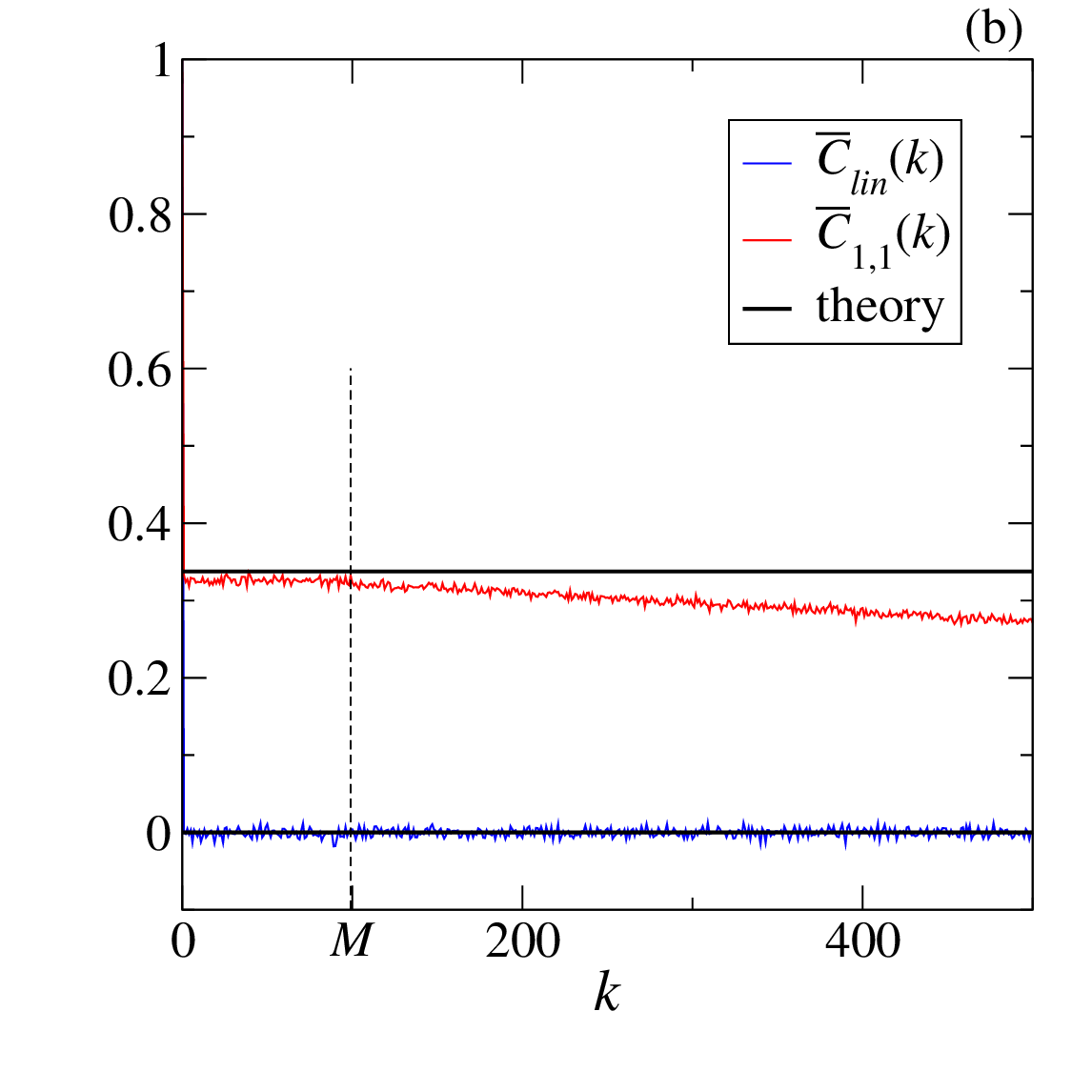}
\caption{
  An autoregressive simulation restores ergodicity up to the time-scale
  $\tau=M\,\Delta t$. Ensemble (a) and time (b) correlations are calculated 
  as in Fig. \ref{fig_prog_corr}.
}
\label{fig_autoregressive}
\end{figure}

In practice, a progressive simulation scheme can be realized by
first extracting a $\sigma$ according to the PDF in 
Eq. (\ref{eq_sigma_cond_ap}), and then an $x_i$ from a Gaussian PDF
with width $\sigma$.
With a different, autoregressive, simulation strategy, scaling and ergodic properties can be
restored together within a good approximation up to a finite time-scale $M$. 
This is obtained by considering a
conditional PDF $p_i^{c,a r}$ which depends, still through
Eq. (\ref{eq_conditional_ap}), on the previous $M-1$ increments only, for
all $i\geq M$:
\begin{equation}
p_i^{c,a r}(x_i|x_{i-1},\ldots,x_{i-M+1})\equiv 
p_M^c(x_i|x_{i-1},\ldots,x_{i-M+1}).
\label{eq_ar}
\end{equation}
After the initial transient of $M$ time-steps, which is realized
according to the progressive scheme in Eqs. (\ref{eq_progressive}), 
using $p_i^{c,a r}$ at each step $i\geq M$ we ``forget'' the increment $x_{i-M}$ and
we thus fix to $M$ the dimension of the conditional PDF for
extracting the next increment of the process. 
This enables the conditional PDF $\rho_M^c$ to wander among all the
ergodic components labeled by the different $\sigma$'s,
as it is shown in Fig. \ref{fig_histo_scaling}a, where we recorded the histogram of the
$\sigma$'s in Eqs. (\ref{eq_conditional_ap},\ref{eq_ar}) 
spanned by both a progressive  and an autoregressive ($M=100$) simulation of 
$10^5$ time-steps. While in the progressive case the histogram is
strongly peaked around a single $\overline\sigma$, in the
autoregressive one it well reproduces the $\rho(\sigma)$ assumed in Eq. (\ref{eq_rho}).  
We define 
the increment over an interval of duration
$\tau=k\,\Delta t$ at time $t=i\,\Delta t$ as 
$Z_{i k}\equiv Y_{i+k}-Y_i$ ($i=1,2,\ldots,N-k$, $k> 0$), 
and then we sample
the the PDF $p_{Z_k}(z)\equiv
\frac{1}{N-k}\sum_{i=1}^{N-k}p_{Z_{i k}}(z)$ 
along a single autoregressive history of $N$ steps.
For an ergodic dynamics it is expected that the scaling properties
of $p_{Z_k}$ reproduce those of $p_{Y_k}$.
Fig. \ref{fig_histo_scaling}b shows that indeed the desired scaling
properties for $p_{Z_k}$,
\begin{equation}
k^D\;p_{Z_k}(k^Dz)=g(z),
\end{equation}
are well satisfied for $D=1/2$ and $k\leq M$.
The fidelity and the ergodicity of the autoregressive simulation are furthermore supported
by an inspection of  $C_{\alpha\beta}(i,j)$ and 
$\overline C_{\alpha\beta}(k)$, which reveals that both the ensemble and the
time correlations approximatively coincide with the theoretical values
as long as $j-i$ and $k$ are smaller than $M$, respectively 
(Fig. \ref{fig_autoregressive}a,b). 
For larger time separations, correlations slowly decay to zero,
producing a smooth crossover to a process with independent increments
on scales much larger than $M$.

For simulations with $D\neq1/2$, 
by considering the rescaled variables $X_1i^\prime\equiv X_i/a_i$, 
with $a_i\equiv\left[i^{2D}-(i-1)^{2D}\right]^{1/2}$ (see main text),  
the above discussion still applies. 
As a consequence, the mechanism of the selection of a specific value 
$\sigma=\overline\sigma$ in $\rho_i^c$ for a single progressive
simulation and that of the dynamical sampling of the various
$\sigma$'s for a single autoregressive one remain valid also when
$D\neq1/2$. 
These features are of crucial importance for the applicability of such kind
of processes in finance \cite{baldovin_3}.

\end{document}